# Improved Performance of Tunneling FET Based on Hetero Graphene Nanoribbons

Fei Liu, Xiaoyan Liu, Jinfeng Kang, Yi Wang

*Abstract*— A heterojunction tunneling field effect transistor based on armchair graphene nanoribbons is proposed and studied using ballistic quantum transport simulation based on 3D real space nonequilibrium Green's function formalism. By using low band gap nanoribbons as the source/drain material, the hetero-structure shows better performance including higher on current, lower off current, and improved steep subthreshold swings compared with homostructure. It is also found the device performance greatly depends on the source/drain dopping density. High doping density leads to fewer density states in the source and degrades the device performance.

*Index Terms*—heterojunction, tunneling transistor, graphene nanoribbon,

## I. INTRODUCTION

Tunneling field-effect transistors(TFET) has attracted intensive research interest for its potential applications for reducing power dissipation in integrated circuits as post-CMOS low-power devices[1, 2]. More recently TFETs based on carbon related materials such as carbon nanotube and armchair graphene nanoribbons (AGNR) have been studied due to the unique properties light effective mass, and direct band gap [3-6]. Theoretical studies show that carbon nanotube (CNT) and graphene nanoribbons can provide rather high $I_{ON}/I_{OFF}$ ratios with subthreshold slopes less than 60 mV/decade. In the AGNR TFET, characteristic of the device is modulated by use of energy band gap by varying the ribbon width [6]. By decreasing the width of ribbon wider energy gap can be get, and $I_{off}$ can be dramatically decreased which results in high $I_{on}/I_{off}$. At the same time, various structures and methods are explored to study the performance of carbon based TFET [7-12]. Strain applied on GNR is used to improve the performance of TFET by modulate the band gap of GNR [8, 9]. TFETs based on GNR heterojunction and partially unzipped carbon nanotubes CNTs are found to have more favorable characteristics for high performance application[10-12].

In this letter, we theoretically study a tunneling FET composed of heterostructure AGNR by using three-dimensional atomistic simulations. Compared to homogenous structure and

This work is supported by NKBRP 2011CBA00604 and 2011ZX02707.
F. Liu, X. Liu, J. Kang and Y. Wang are with the Key Laboratory of Microelectronic Devices and Circuits, Institute of Microelectronics, Peking University, Beijing 100871, China (e-mail: xyliu@ime.pku.edu.cn).
Digital Object Identifier

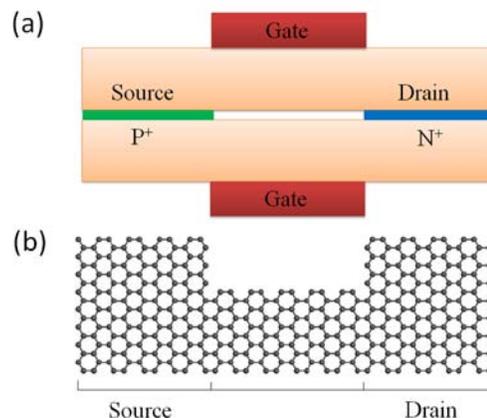

Fig. 1 (a) Schematic view of double gate GNR TFETs with $SiO_2$ oxide thickness of $t_{ox} = 1.0$ nm and dielectric constant of k=4 (b) Atomistic structure of GNR heterostructure, which is composed of N=16 AGNR in source/ drain region and N=10 AGNR in the channel . The p-doped source and n-doped drain have the same doping density of 0.005 dopant/atom. Source and drain have the same length of 10 nm and channel length is 16nm.

previously studied heterostructure GNR tunneling FET, the proposed heterostructure TFET uses wide ribbon with small band gap in the source extension. Thus a reduced tunneling barrier at the on state can be set up, which is helpful to enhance on current; whereas large band gap ribbon is used as channel material to keep large energy barrier at the off state and get low off current. Our results demonstrate that high $I_{on}/I_{off}$ and much smaller subthreshold slopes can be achieved in the heterostructure TFET. By modifying source/drain doping, the device performance can be modulated. It is observed that hetero TFET with low source/drain doping exhibits a better performance in terms of $I_{ON}/I_{OFF}$ and steep subthreshold swings.

## II. SIMULATION APPROACH

As illustrated schematically in Fig. 1 (a), the modeled p-i-n tunneling FETs has a double-gate structure with a gate oxide thickness of 1.0 nm and the dielectric constant of κ= 4 for $SiO_2$. N1 and N2 denote the number of atoms in transverse direction in source/drain region and gate channel region, respectively. Because the band gap of AGNR depends on the width of ribbon, we can design heterojunction by use of different width ribbons. For N1=N2, the whole transport part of TFET has the same width and is homogenous; otherwise, the whole channel is a heterojunction composed of two kinds of semiconducting AGNR with different band gap. Source and drain with wider AGNR than that of the channel is studied here as shown in Fig.

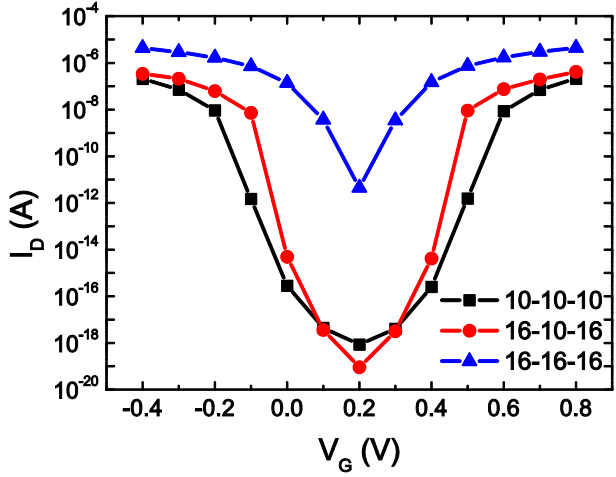

Fig. 2 $I_D$–$V_G$ characteristics of GNR TFETs with different atomic junction structures at $V_D = 0.4$ V with dopping density of $N_{S/D}=5\times 10^{-3}$ dopant /atom.. 10-10-10 and 16-16-16 denote homogenous junction with the atom number N=10 and N=16 in transverse direction. 16-10-16 represents hetero-structure with width of source /drain N1=16 and width of channel N2=10.

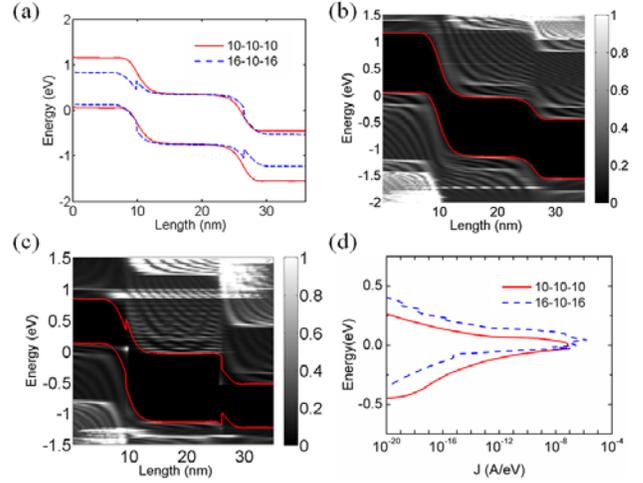

Fig. 3 (a) Band profiles of 10-10-10 homojunction and 16-10-16 hetero-junction at off state at $V_G$=0.2V and $V_D$=0.4V. (b) and (c) shows local density of states of the homojunction and the heterojunction at on state at $V_G$=0.6V and $V_D$=0.4V . (d) the compared energy resolved current spectrum at on state of two structures.

1(b) with N1=16 and N2=10. N1-N2-N1 is used to denote the junction structure.

The device characteristics in ballistic limit are studied by self-consistently solving the 3D Poisson equation and open boundary schrodinger equation using the non-equilibrium Green's function formalism [13]. The Hamiltonian of the device is described using nearest-neighbor tight binding (TB) approximation with a single $p_z$ orbital and a hopping energy $t_0$=-2.7 eV. In considering edge-bond relaxation, the hopping energy along the edges of the ribbon is modified as t = 1.02 $t_0$.

### III. RESULTS AND DISCUSSION

Transfer characteristics of AGNR TFET with different channel configurations are shown in Fig. 2. Two structures are composed of homogenous AGNR along the channel with N1=N2=10 and N1=N2=16 respectively, and another structure is heterojunction with N1=10 and N2=16. From Fig. 2, we can see that the performance of homojunction greatly depend on the width of ribbons. With the width increasing from N1=10 to N1=16, $I_{ON}$ and $I_{OFF}$ increase while $I_{on}/I_{off}$ decreases greatly. The degrading of device performance results from the change of band gap with width variation of GNR. For armchair GNR the band gap is determined by the width, which is in general inversely proportional to the GNR width. As the GNR width decreases from N1=16 to N1=10 , the band-gap increases from 0.70 eV to 1.10 eV, leading to smaller on and off current due to the suppressed tunneling effect. Larger band gap usually leads to larger $I_{ON}/I_{OFF}$ ratio but low $I_{ON}$. In order to maintain lower $I_{OFF}$ and to get higher $I_{ON}$, we use wider ribbon to substitute the source and drain of 10-10-10 and have 16-10-16 heterojunction as transport material. Compared to 10-10-10 homogenous structure, the $I_{OFF}$ of 16-10-16 hetero-junction get smaller by one order while $I_{ON}$ at $V_G$=0.6 V increases by one order and lies between those of two homogenous structure, which means higher $I_{ON}/I_{OFF}$. Another interesting feature of heterojuction is that it has much smaller subthreshold slopes (SS) down to 27.29 mV/dec than 47.96 mV/dec of N1=10 homostructure AGNR TFET.

In a tunneling FET source-drain current is controlled by modulating the tunneling barrier from the source valence band to channel conduction band. At off state, the source valence band is below channel conduction band. Hence direct tunneling from source to drain (DTSD) plays an important role. Fig 3 (a) show the conduction and valance band profiles of 10-10-10 homojunction and 16-10-16 heterojunction at off state under the gate voltage $V_G$ =0.2 V. We can see there is nearly the same tunneling barrier across the channel while a little difference appears in the channel drain junction. The tunneling barrier difference is due to less voltage drop in the channel of heterostructure. Thus, hetero TFET has a larger tunneling barrier in the region which leads to smaller off current. At on state, the current mainly relies on band to band tunneling (BTBT) from the source valence band to the channel conduction band. Fig. 3 (b) and (c) show the local density of states and band profiles of 10-10-10 homogenous channel and 16-10-16 hetero-junction at $V_G$=0.6 V and $V_D$=0.4 V. It can be seen that under the voltage and the doping density channel conduction bands of both devices are suppressed below source valence bands. So there is an energy window for BTBT from source valance band to channel conduction band, which dominates the current at on state. Due to different band gap of the two structures the energy window between source valence band and channel conduction band of 16-10-16 heterojunction is wider than that of 10-10-10 homojunction; moreover, heterojunction has a thinner tunneling barrier. Fig. 3 (d) shows a comparison of the energy resolved current spectrum for two geometries at the on state, which confirms the aforementioned phenomena. It is shown that 16-10-16 hetero structure has a larger BTBT energy window and larger BTBT current. Below the energy of channel conduction band, 10-10-10 homostructure has a larger DTSD current due to the thinner barrier in the source channel junction, which is similar as the case at off state.

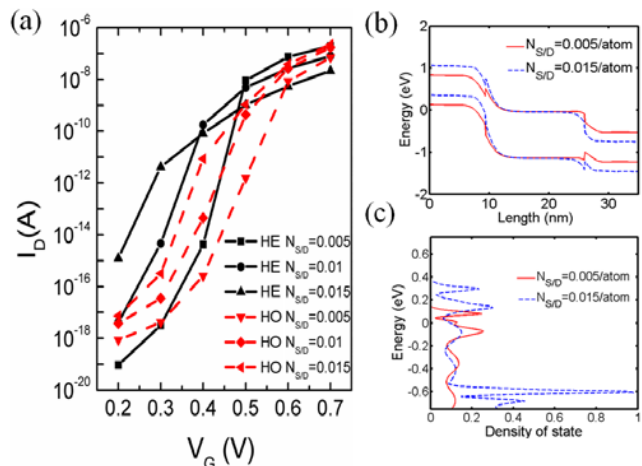

Fig. 3 (a) $I_D$–$V_G$ characteristics of 16-10-16 hetero (HE) and 10-10-10 homo (HO) GNR TFETs with different doping density of $N_{S/D}$=5×10$^{-3}$ , 1×10$^{-2}$ and 1.5×10$^{-2}$ dopant /atom.. (b) band profiles  and (c) local density of states of 16-10-16 hetero-stucture at the on state with doping density of $N_{S/D}$=5×10$^{-3}$ and 1.5×10$^{-2}$ dopant /atom..

Compared to BTBT current, DTSD current is much smaller. Consequently, 16-10-16 heterojunction has an improved on current for the contribution of BTBT effect.

Next we focus on source/drain doping effect on device performance for a given device structure. Fig. 4 shows a comparison $I_D$-$V_G$ characteristics for 16-10-16 hetero AGNR TFET and 10-10-10 homostructure with different source/drain doping density of $N_{S/D}$=5×10$^{-3}$ , 1×10$^{-2}$ and 1.5×10$^{-2}$ dopant /atom. The figure shows that device performance greatly depends on the doping concentrations. For homostructure, current increases significantly for all gate voltages with the increase of source/drain doping density as shown in Fig. 3 (a). This is due to the fact that more carriers are available at the source region and the direct tunneling thickness from source to channel gets narrower as $N_{S/D}$ gets higher [14]. While, the situation of heterostructure is different from the case of homostructure. With the increase of $N_{S/D}$, the current at a high gate voltage $V_G$>0.5V gets smaller. In order to study this phenomenon, band diagram and local density of state at the source center of 16-10-16 hetero-structure under $V_G$=0.6V are presented in Fig.3 (b) and (c), respectively. Fig.3 (b) shows that a higher doping density decreases the tunneling barrier and increases the band to band tunneling at the source channel junction. However, the thinner tunneling barrier doesn't enlarge the current. This is because the tunneling barrier is not the only aspect determining the on current. Available carriers at the source also play an important role. Local density of state at the center of sources is shown in the Fig. 3 (c). We can see that there are fewer states at the source around the fermi energy due to the increase of doping density which leads to lower $I_{ON}$, even though there is thinner  tunneling barrier.

## IV. Conclusion

In this paper, we perform simulation study on the heterostructure tunneling field transistor composed of two deferent width graphene nanoribbon. In the hetero-structure tunneling FET, wide ribbon with small band gap is used as source material, which results in reduced tunneling barrier at the source junction and leads to high on current.  At the same time, the large band gap ribbon in the channel helps to get low off current. Consequently, the performance of the hetero TFET is greatly improved with higher $I_{ON}/I_{OFF}$ and smaller subthreshold slope. Besides, we analyze doping effect on the device performance and find that even though higher doping density will make the tunneling barrier at the source-channel junction thinner at on state, fewer available electron states around Fermi energy will finally leads to smaller on current and deteriorates device characteristics. These results may be helpful to design heterostructure tunneling FET.

## References

[1] A. Seabaugh and Q. Zhang, "Low-voltage tunnel transistors for beyond-CMOS logic," *Proc. IEEE*, vol. 98, no. 12, pp. 2095–2110, Dec. 2010.
[2] A. M. Ionescu and H. Riel, "Tunnel field-effect transistors as energy efficient electronic switches," *Nature*, vol. 479, no. 7373, pp. 329–337, Nov. 2011.
[3] J. Appenzeller, Y.-M. Lin, J. Knoch, and Ph. Avouris, "Band-to-band tunneling in carbon nanotube field-effect transistors", *Phys. Rev. Lett*. Vol. 93, p. 196805, 2004.
[4] S. O. Koswatta,M. S. Lundstrom, and D. E. Nikonov, "Band-to-band tunneling in a carbon nanotube metal-oxide-semiconductor field-effect transistor is dominated by phonon-assisted tunneling", *Nano Lett.* Vol.7, No.5, p.1160-1164, 2007.
[5] Q. Zhang, T. Fang, H. Xing, A. Seabaugh, and D. Jena, "Graphene Nanoribbon Tunnel Transistors", *IEEE Electron Device Lett.*, vol. 29,no. 12, pp. 1344–1346, Dec. 2008.
[6] P. Zhao, J. Chauhan, and J. Guo, "Computational study of tunneling transistor based on graphene nanoribbon", *Nano Lett.* 9, 684, 2009.
[7] G. Fiori and G. Iannaccone, " Ultralow-voltage bilayer graphene tunnel FET", *IEEE Electron Device Lett.*, vol. 30, no. 10, pp. 1096–1098,Oct. 2009.
[8] Y. Lu, J. Guo, "Local strain in tunneling transistors based on graphene nanoribbons", *Appl. Phys. Lett.* vol. 97, p. 073105 , 2010
[9] J. H. Kang, Y. He, J. Y. Zhang, X. X. Yu, X. M. Guan, and Z. P. Yu, "Modeling and simulation of uniaxial strain effects in armchair graphene nanoribbon tunneling field effect transistors s", *Appl. Phys. Lett.* vol. 96, p. 252105, 2010
[10] K. T. Lam, D. W. Seah, S. K. Chin, S. B. Kumar, G. Samudra, Y. C. Yeo, and G. Liang, " A Simulation Study of Graphene nanoribbon tunneling FET with heterojunction channel", **IEEE Electron Device Lett.**, vol. 31, no. 6, pp. 555–557, Dec. 2008.
[11] Y. K. Yoon and S. Salahuddinb, "Barrier-free tunneling in a carbon heterojunction transistor", *Appl. Phys. Lett.* vol. 97, p. 033102 , 2010
[12] Y. K. Yoon, S. H. Kim, and S. Salahuddina,"Performance analysis of carbon-based tunnel field-effect transistors for high frequency and ultralow power applications", *Appl. Phys. Lett.* vol. 97, p. 233504 , 2010
[13] S. Datta, "Quantum Transport: Atom to Transistor", Cambridge University Press, Cambridge, 2005.
[14] S. K. Chin, D. W. Seah, K. T. Lam, G. S. Samudra,and G. Liang, "Device Physics and Characteristics of Graphene Nanoribbon Tunneling FETs", *IEEE Trans. Electron Devices,* vol. 57, no. 11, p. 3144-3152, 2010